\documentclass[12pt,preprint]{aastex}

\slugcomment{Submitted to ApJ June 7, accepted March 17, 2007}

\shorttitle{Accurate Mass Determination of AB Dor C with Theoretical Evolutionary Tracks}
\shortauthors{Close et al.}

\begin{document}


\title{New Photometry and Spectra of AB Doradus C: An Accurate Mass Determination of a Young Low-Mass Object with Theoretical Evolutionary Tracks{\footnotemark}}

\footnotetext{ Based on observations made with ESO Telescopes at the Paranal Observatories under programme ID 276.C-5013.}


\author{Laird M. Close\altaffilmark{2}, Niranjan Thatte\altaffilmark{3}, Eric L. Nielsen\altaffilmark{2}, Roberto Abuter\altaffilmark{4}, Fraser Clarke\altaffilmark{3}, \& Matthias Tecza\altaffilmark{3}}

\email{lclose@as.arizona.edu}

\altaffiltext{2}{Steward Observatory, University of Arizona, Tucson, AZ 85721}
\altaffiltext{3}{Dept. of Astrophysics, DWB, University of Oxford, Keble Road, 
Oxford OX1 3RH, U.K.}
%\affil{$^2$Institute for Astronomy, University of Hawaii, Honolulu, HI}
\altaffiltext{4}{European Southern Observatory, Garching, Germany}   



\begin{abstract} 

We present new photometric and spectroscopic measurements for the
unique, young, low-mass evolutionary track calibrator AB Dor C. While
the new Ks photometry is similar to that previously published in Close
et al. (2005) the spectral type is found to be earlier.  Based on
new H \& K IFS spectra of AB Dor C (Thatte et al. 2007; paper I) we
adopt a spectral type of M$5.5\pm1.0$ for AB Dor C. This is
considerably earlier than the M$8\pm1$ estimated in Close et
al. (2005) and Nielsen et al. (2005) yet is consistent with the
M$6\pm1$ independently derived by Luhman \& Potter (2005). However,
the spectrum presented in paper 1 and analyzed here is a significant
improvement over any previous spectrum of AB Dor C. We also present
new astrometry for the system which further supports a $0.090\pm0.005
M_{\sun}$ mass for the system. Once armed with an accurate spectrum
and Ks flux we find $L=0.0021\pm0.0005 L_{\sun}$ and
$T_{eff}=2925^{+170}_{-145}$K for AB Dor C. These values are
consistent with a $\sim75$ Myr $0.090\pm0.005 M_{\sun}$ object like AB
Dor C according to the DUSTY evolutionary tracks (Chabrier et
al. 2000). Hence masses can be estimated from the HR diagram with the
DUSTY tracks for young low-mass objects like AB Dor C.  However, we
cautiously note that underestimates of the mass from the tracks can
occur if one lacks a proper (continuum preserved) spectra or is
relying on NIR fluxes alone.

\end{abstract}


\section {Introduction}

There is currently great interest in the direct detection of
extra-solar planets, brown dwarfs, and very low-mass stars.  The study
of such young, low-mass objects has been yielding increasingly
fruitful science, yet the field remains dependent on evolutionary
models to properly interpret the data that are collected from these
objects.  In particular, mass, while a fundamental property, is very
rarely measured directly (through orbital dynamics), and instead must
be inferred from theoretical tracks (e.g., Burrows, Sudarsky, and
Lunine 2003, Chabrier et al. 2000).  It is thus of great interest to
discover calibrating objects that can link a dynamically measured mass
with observables such as accurate NIR (1-2.5 $\mu$m) fluxes and
spectral types.

AB Dor A was suspected of having a low-mass companion due to VLBI and
Hipparcos measurements of an astrometric wobble (Guirado et
al. 1997). Recently Close et al. (2005), reported the first direct
detection of this low-mass companion (AB Dor C) to the young star AB
Dor A, along with the first measurements of the JHKs fluxes, spectral type, and
dynamically determined mass of AB Dor C.  Upon comparing these results
with the predictions of Chabrier et al. (2000), they found the models
to be systematically over-predicting the fluxes and temperature of AB
Dor C (especially at J \& H), given a system age of 50 Myr.  Put
another way, the model masses seem to be underestimating the mass of a
low-mass object given its age, J \& H NIR fluxes, and spectral type.
Since the publication of these results, another calibrating object has
been reported by Reiners, Basri, and Mohanty (2005): USco CTIO 5.
While this equal-mass binary is younger ($\sim$8 Myr) and more massive
(total mass $\ge$0.64 M$_{\sun}$) than AB Dor C, Reiners et al. find
the same trend of models under-predicting masses based simply on
photometric and spectroscopic data applied to the HR diagram.  A
similar trend for such masses was previously noted by Hillenbrand and
White (2004).  Moreover, this trend has been theoretically predicted
for higher masses by Mohanty, Jayawardhana, and Basri (2004), and by
Marley et al. (2005) for planetary masses. Hence it is critically
important to accurately calibrate the evolutionary tracks to determine
if there are systematic errors. In particular, obtaining an accurate $T_{eff}$ from spectra of low-mass, young, objects is challenging and makes comparison to evolutionary tracks difficult (Chabrier et al. 2005).

Recently Nielsen et al. (2005) have obtained new orbital epochs and
confirmed the previous mass of AB Dor C of $0.090\pm0.005 M_{\sun}$ with the technique of 
Guirado et al. (2006). Hence AB Dor C has
a uniquely well known dynamical mass for a young ($\sim 75$ Myr)
low-mass object. Only GL 569Ba/Bb has a similarly well determined mass
but is thought to be somewhat older (age $\sim$300 Myr; Zapatero
Osorio et al. 2005), though a younger age ($\sim 100$ Myr) and
binarity of GL 569 Ba has recently been proposed (Simon et
al. 2006). Also the lack of lithium in the spectrum of GL 569B
(Zapatero Osorio et al. 2006) is surprising. Hence, there is a need to
quantify the AB Dor A/C system as closely as possible to have an
additional calibrator for low-mass young evolutionary models.

The age of AB Dor (and its associated moving group) is also somewhat
uncertain, Luhman et al. (2005) argue for an older age of the AB Dor
group of 75-150 Myr while Nielsen et al.(2005) find $75\pm25$
Myr. Recently Lopez-Santiago et al. (2006) argue for a 50 Myr age for
the core of the AB Dor moving group (which includes AB Dor A) as did Zuckerman et al. (2004) in the
original ``AB Dor moving group'' paper. Recently Janson et al. (2007) find an age range of 50-100 Myr. Here we adopt an ``average'' age of 75 Myr for the system.

There is also uncertainty in the spectral type of AB Dor C. The small
separation between A \& C of only $0.155\arcsec$ combined with the
$>100$ contrast made an accurate spectral type difficult to measure in
the dataset of Close et al. (2005). A reanalysis of the Close et
al. spectra by Nielsen et al. (2005) suggested a spectral type of
$M8\pm1$. However, an independent re-analysis of these data by Luhman
\& Potter (2005) suggested a spectral type of $M6\pm1$. The importance
of this temperature is paramount to plotting AB Dor C on the HR
diagram to calibrate the accuracy of the evolutionary tracks. These
past papers highlight the difficulty in trying to determine the
spectral type of a faint companion within $0.16\arcsec$ of a bright
star with an AO-fed long-slit spectrograph. Indeed none of these past
reductions of the AO long-slit spectra were able to preserve the
continuum of AB Dor C. Hence the true spectrum of C could not be
accurately determined until now.

In the companion paper to this one (Thatte et al. 2007; hereafter
Paper I) we report on new, excellent integral field spectroscopy (IFS) of
AB Dor A \& C, observed with the SINFONI instrument at the VLT.
Using the new data analysis technique of PSF scaling
and differencing (PSD), we were able to effectively eliminate all
contamination from AB Dor A, to produce a very high signal-to-noise
spectrum of AB Dor C, which also preserves the continuum. The PSD
technique is able to achieve very high contrast ($\sim$9 mags at
0\farcs 2) without a coronagraph, and without any prior assumptions
about the spectral characteristics of the companion object (see Paper 1 for more inforamtion about the IFS reduction, and the PSD technique).

In this paper we present new VLT science archive Ks photometry and astrometry
for AB Dor C and a more accurate H \& K spectrum and a new temperature of AB Dor C. Then we compare the
accuracy of the mass of AB Dor C predicted by the popular DUSTY
theoretical tracks (Chabrier et al. 2000).

\section{OBSERVATIONS \& REDUCTIONS}

The AB Dor system (A, C, and their companion BaBb -- $8.87\arcsec$ distant) was observed
with the VLT NACO AO system (Lenzen et al. 2003; Rousset et al. 2002)
on Jan 7, 2005 by R. Neuhauser et al. in the Ks band. We have
reduced these archive data with our NIR AO data reduction IRAF pipeline
(Close et al. 2002, 2003).
 
The Ks images were fully flat-fielded, sky \& dark subtracted, bad
pixel cleaned, and aligned and medianed (see Close et al. 2003 for
more details about our AO pipeline). Only the first 18 200x0.347 second Ks
images were reduced since the seeing became worse after
these first 18 images.

We also observed AB Dor A \& C with the SINFONI IFS as described in paper I. See paper I for extensive details of the IFS observations \& reductions.

\section{ANALYSIS}

\subsection{The Ks Flux and Astrometry of AB Dor C}

To accurately measure the flux of a tight faint companion is never
trivial. However, by January 2005 AB Dor C had moved out to a
$0.22\arcsec$ separation from AB Dor A. While this may seem a very small
separation it is considerably better than the $0.155\arcsec$ separation
during the Feb 2004 discovery images of Close et al. (2005). Hence, we
should be able to better gauge the flux of AB Dor C at this latter epoch.

To measure the brightness of AB Dor C we utilized the unsaturated PSF
image (in the narrow band 2.12 micron filter) of AB Dor A which was
taken just before the AB Dor Ks data set. This 2.12 $\mu m$ ``PSF'' image
was taken at the same airmass and seeing as the Ks images. Also, the
exposure times were the same (DIT=0.347 seconds, NDIT=200 coadds) for
the PSF and Ks images. The FWHM of the PSF appeared similar to that of
AB Dor C. So we have some confidence that this was a good PSF.

We shifted (with a cubic spline) and scaled this PSF image until
subtracting it led to the flux from C being completely removed from
the reduced Ks images. However, as is clear from Figure 1, there is
some residual flux or ``super-speckles'' surrounding C's
position from A. Therefore, it is impossible to absolutely determine the flux
of C due to some uncertainty in A's PSF. We adopt the mean flux of AB
Dor C as that where the residual flux of A at the position of C is
equal to that 180 degrees on the other side of the A PSF (see bottom
left image in figure 1). This is a reasonable assumption since the
super speckles in the NACO PSF are mainly due to phase errors which
transform to symmetric pairs on either side of the PSF.
Here we have made the conservative assumption that
the possible range of C's flux should be from completely removing
all residual flux at the position of C (Fig 1: lower right) to assuming there is a slightly brighter
super-speckle at C's position (Fig 1; upper right).

\begin{figure}
 \includegraphics[angle=0,width=\columnwidth]{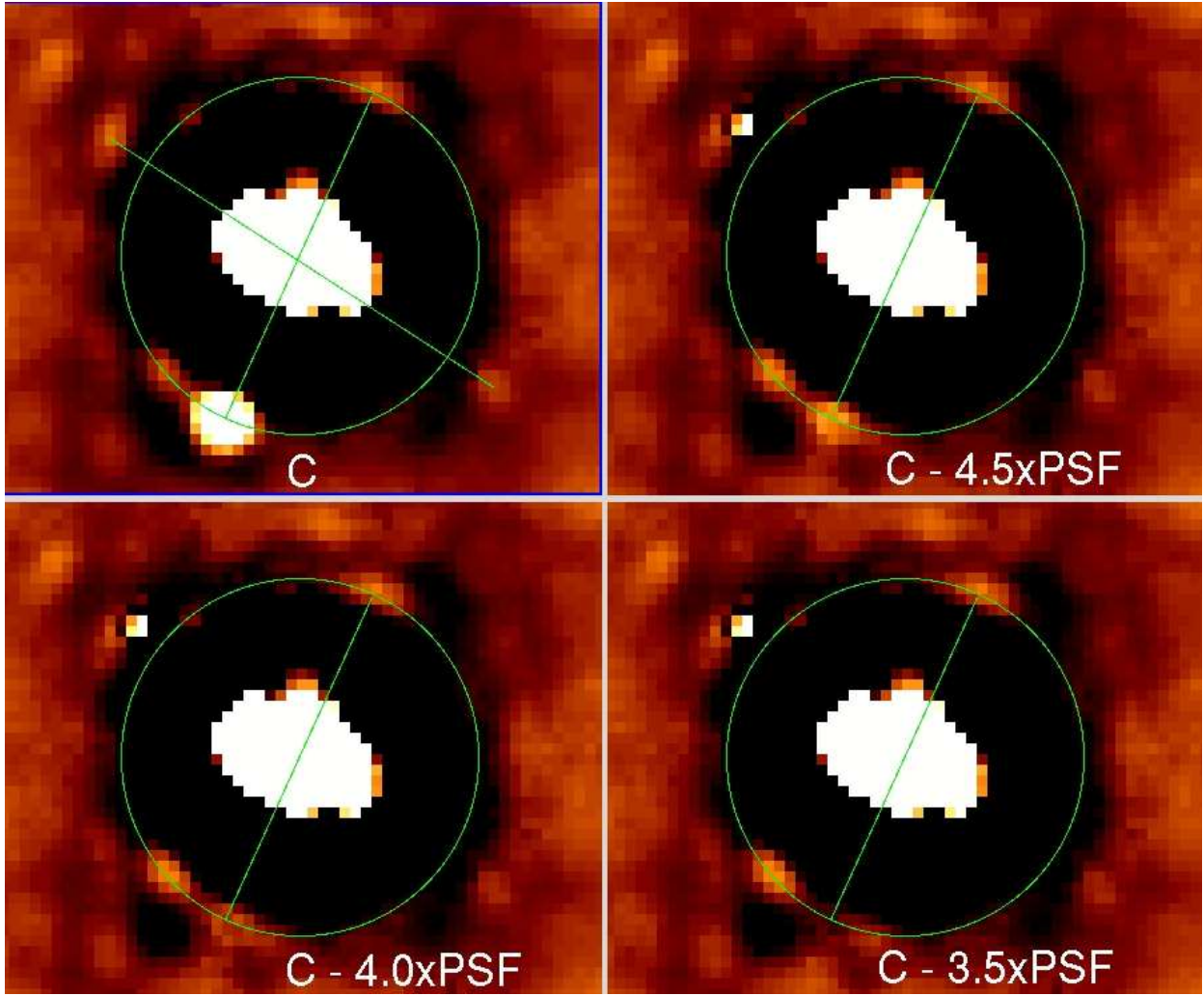}
\caption{
Top left: AB Dor C and A (unsharp masked). Top right: After subtracting the PSF scaled by 4.5x (Ks=9.66 mag) there is too much residual light at the location of C -- hence this is a lower limit to the flux of AB Dor C. Bottom left: A more optimal subtraction leaving the same amount of flux on either side of A (Ks=9.50 mag). Bottom right: An over-subtraction of the PSF leading to an upper-limit of the flux of C of Ks=9.34. The lines bisect the AB Dor A PSF and mark the location of AB Dor C.  
}
\label{fig1}
\end{figure}

%EDITOR place FIGURE 1 HERE
\placefigure{fig1} 

\subsubsection{Photometry of AB Dor C}

The Ks flux of AB Dor B is known from the 2MASS Catalog to be
$Ks_B=7.435$ (corrected for the 0.095 mag of contamination of B due to
A in the 2MASS $4\arcsec$ aperture photometry). Hence by calibrating
the mean delta Ks of the PSF to BaBb (which varied by $<2$\% over a
range of 1.5,2.0,\& 2.5$\arcsec$ apertures) we derive a
$Ks=9.50\pm0.16$ mag for C. This flux is derived assuming a symmetric
PSF while the large errorbars are derived assuming minimal symmetry in
the AB Dor A PSF and adding in the maximum errors of the aperture
photometry (see the top right and bottom right images in Figure
1). Hence, the true flux of AB Dor C should certainly fall inside this
range. Our new value of $Ks=9.50\pm0.16$ is very similar to the
$9.45^{+0.060}_{-0.075}$ mag measured by Close et al. (2005). Our
conservative assumption of minimal PSF symmetry around AB Dor A leads
to our photometric errors being larger than those of Close et al. who
assumed the PSF was mainly symmetric. Our new photometry is also
consistent (if a bit brighter and more precise) than the
$9.79^{+0.25}_{-0.33}$ mag independently determined by Luhman \&
Potter (2005) from the Close et al.(2005) dataset. Moreover, we obtain
a $Ks\sim9.59$ mag from the IFS data-cube in paper 1 -- adding further
confidence to our Ks band flux of C. Hence, there is now reasonable
agreement from three different epochs that $Ks=9.5\pm0.16$ for AB Dor C.

From the Hipparcos distance of $14.9\pm0.1$ pc and the $-2.82\pm0.15$ mag $BC_K$ of Allen et al. (2003)(appropriate for M5.5) and noting that for an M5.5 $Ks-K\sim 0.03$ mag (Daemgen et al 2006) we derive a luminosity of $L=0.0021\pm0.0005 L_{\sun}$ from the observed $Ks=9.50\pm0.16$ mag of AB Dor C.

\subsubsection{Astrometry of the AB Dor A/C system}

From our PSF fitting AB Dor C is found to be
$0.219\pm 0.008\arcsec$ at $PA=155.92\pm0.5^{o}$ in good agreement with
the orbit of Nielsen et al. (2005) which predicts a separation of
$0.2186\arcsec$ at $PA=155.175^{o}$ for Jan 7, 2005 (2005.0170). This good
agreement gives us further confidence in the orbital solution of
Nielsen et al. (2005). Moreover, we have another (even later) Jan 24 2006 (2006.0660) epoch of
observation from the IFS dataset of Paper I (see Table 1). This additional data-point
also fits the orbital solution of Nielsen et al. (2005) very
well. Figures 2 \& 3 illustrate the quality of the orbital solution of
Nielsen et al. (2005). Note that this orbit was mainly determined by
the reflex motion of A from $\sim1$ mas Hipparcos/VLBI astrometry of
Guirado et al. (2006) (for clarity these Hipparcos/VLBI data-points are
not shown in Figure 2). 

Recently Jeffers, Donati \& Cameron (2007) have estimated the radial
velocity variations of the AB Dor A/C system from 1988 to 1994. They
find a reasonable fit of our older Close et al. (2005) orbital
solution to their radial velocity measurements. The more recent
orbital solution of Nielsen et al. (2005) should be similar. This
further proof of the quality of the A/C orbital solution.

\placetable{tbl-1}

\subsubsection {Astrometry of the Ba Bb system}

We were also able to study the companion binary BaBb (a tight
$0.07\arcsec$, $PA=238.6\pm0.3$ binary discovered by Close et
al. 2005) some $9\arcsec$ away is visible in the reduced image. On Jan
7, 2005 we find the BaBb binary is now $0.060\pm 0.003\arcsec$ in
separation at $PA=246\pm 2$ (with $\Delta K_{s}$=0.22 mag), while Ba
is $8.87\pm 0.10\arcsec$ at $PA=346.31\pm0.5$ w.r.t. A (see Table 1
for more details). The small change in position of Bb w.r.t. Ba \& A
since the Feb. 2, 2004 Close et al. (2005) observations definitively
proves BaBb is a tight binary itself bound to A. Previous observations show that B itself is in orbit around A (Guirado et al. 2006).

Although we know relatively little about the orbit of Bb around Ba we know 
AB Dor Ba/Bb has a projected separation of 0.9 AU, and for an almost equal 
mass binary with combined spectral type $\sim$M4 (Martin \& Brandner 1995; see Table 1), this 
is consistent with a Keplerian orbit with a period of order one year.  We 
have also obtained an aquisition image from the ESO VLT archive between 
the two observational epochs discussed here.  This image (from ESO program 
073.C-0469(B), Chauvin et al.) was taken on September 24, 2004: about 
seven and a half months after the Close et al. 2005 dataset, and about 
three and a half months prior to the Neuhaser et al. dataset.  In this 
image, taken with the same NACO system, using the same S13 camera with 
identical platescale, the same Ks filter, and AB Dor Ba/Bb having a 
similar FWHM compared to future and past epochs, the appearance of the 
binary is drastically different.  Unlike the Feb. 2004 and Jan. 2005 
images, where the two components of the binary are clearly 
distinguishable, the Sep. 2004 image appears to be of a single star.  This 
short-period variation, combined with the similar separations and 
position angles observed between Feb. 2004 and Jan. 2005 seem to constrain 
the orbit of the AB Dor Ba/Bb system to approximately 11 months.  Future 
monitoring of this system is required to obtain more detailed orbital 
parameters.

\begin{figure}
 \includegraphics[angle=0,width=\columnwidth]{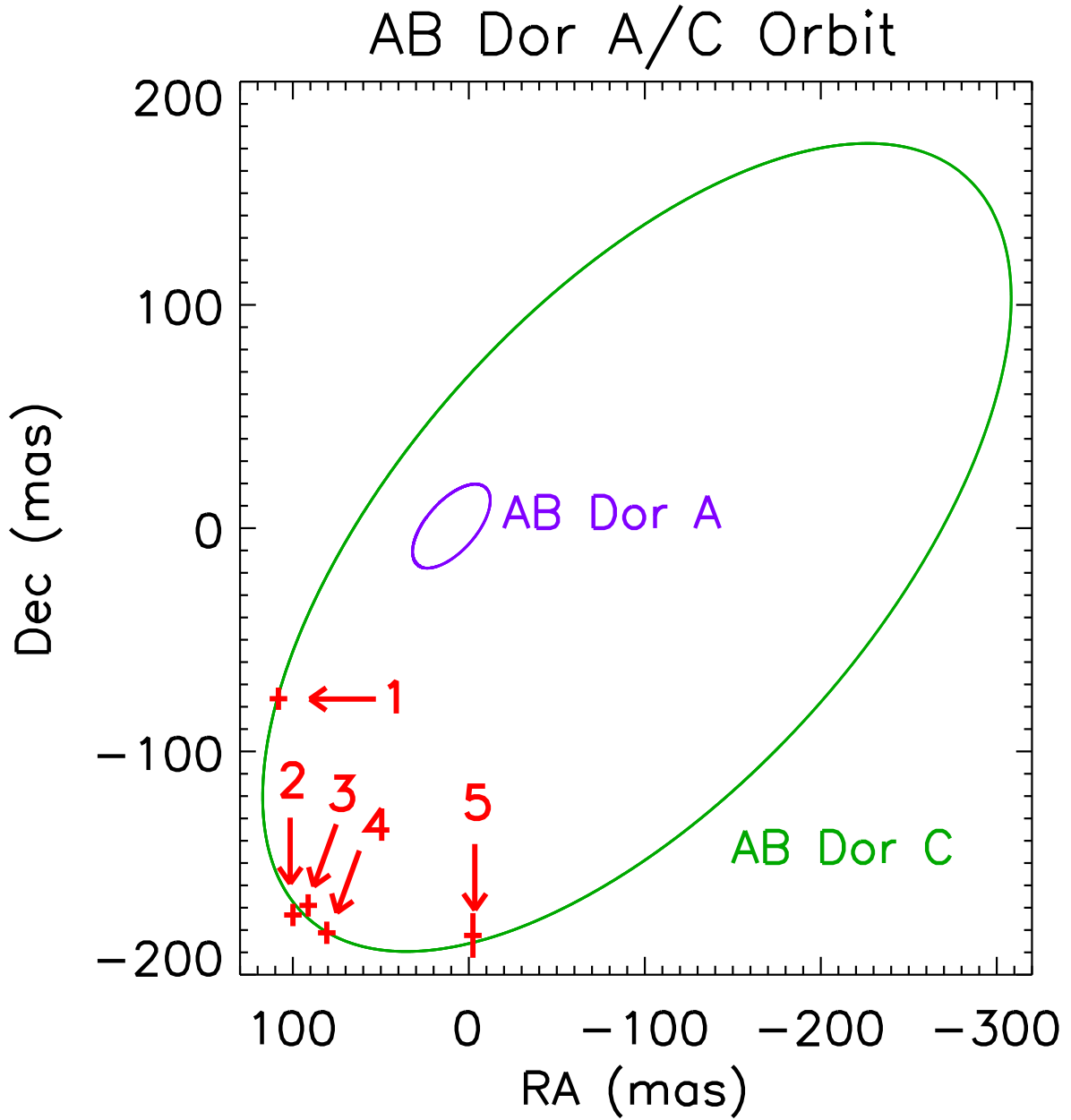}
\caption{
The orbital solution of Nielsen et al. (2005) with our 2 new data points (\#4 this work \& \#5 paper 1) added. Note that this orbit was mainly determined by the reflex motion of A from $\sim1$ mas Hipparcos/VLBI astrometry of Guirado et al. (2006) (for clarity these Hipparcos/VLBI datapoints of A's motion are not shown above -- see Guirado et al. (2006) for details).  
}
\label{fig2}
\end{figure}

\begin{figure}
 \includegraphics[angle=0,width=\columnwidth]{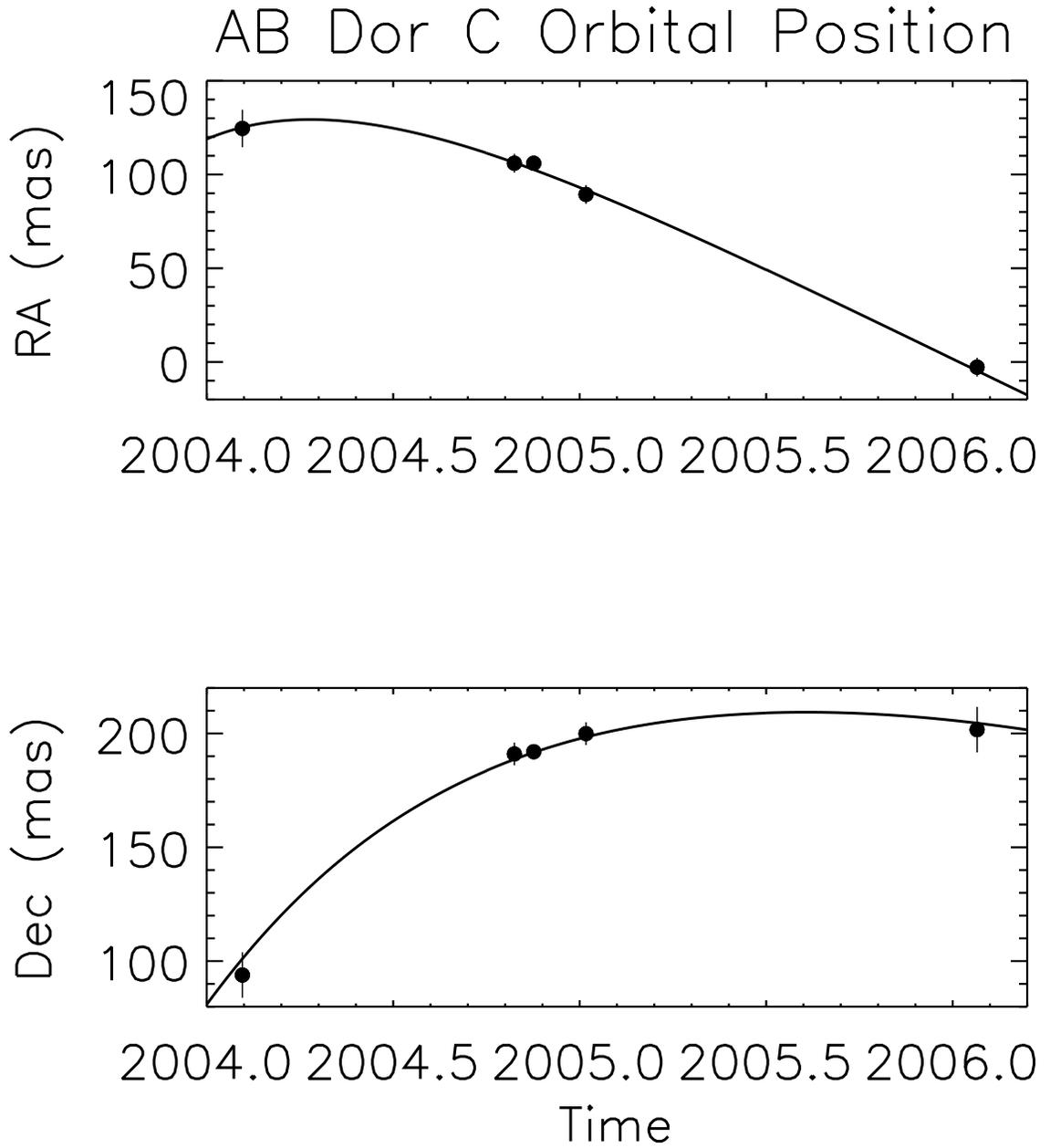}
\caption{ The separation of AB Dor C w.r.t. A as a function of time. The
solid line is the orbital solution of Nielsen et al. (2005) with our
new data points (the most recent two datapoints) added on top of the predicted orbit (this is not a new fit). The agreement
is excellent with both the new Ks and IFS astrometric
datapoints. Hence the mass of AB Dor C is confirmed to be within the
range $0.090\pm0.005 M_{\sun}$.  }
\label{fig3}
\end{figure}

\placefigure{fig2}
\placefigure{fig3}

\subsection{The Spectral Type of AB Dor C}

As is outlined in paper I an excellent R$\sim$1500 H and Ks spectrum of AB Dor
C was obtained with the SINFONI IFS. Here we attempt to place our spectrum
on a spectral sequence. Unfortunately there simply are no good H \& K
template spectra for late M stars of $\sim$75 Myr of age. Hence, we have
endeavored to fit our spectrum to young ($\sim$ 1 Myr; Gorlova et
al. 2003) and old field M stars (Cushing et al. 2005). As is clear
from figures 4 - 7 there is pretty good agreement that the spectrum is
similar to M5 or a M6.

   We feel the spectra presented here are superior to those previously
   published by Close et al. (2005); Nielsen et al. (2005); and Luhman
   \& Potter (2006) for several reasons. All previous published
   spectra were different reductions of the long-slit NACO K spectra
   obtained by Close et al. (2005). These AO-fed long slit spectra
   suffered from many of the drawbacks of AO-fed long-slit
   spectra. First, the very small $0.085\arcsec$ slit used in the
   older spectra was only roughly aligned with the binary, due to
   flexure etc. Hence, there are different slit losses for each
   spectrum obtained as AB Dor was nodded along the slit. Moreover, as
   the core FWHM of the AO PSF decreases as $\lambda$ increases, hence
   slit losses are also function of the wavelength as well as the
   centering error. Finally there may have been some error in the PA
   angle of the slit and so there may be different slit losses between
   AB Dor A and C. Also the very small separation of $0.15\arcsec$
   between A \& C made these spectra very difficult to reduce (in fact
   many of the individual spectra had to be dropped from the reduction
   since no manner of subtracting A could reveal a non-noise signal at
   the position of C). Indeed, Luhman \& Potter (2006) and Nielsen et
   al. (2005) only use $\sim$ 50\% of the spectra obtained to try and
   detect a signal from C. In all these previous spectra the flux from
   AB Dor A swamped the light from C (just 0.15\arcsec away); this is
   not surprising given the $\sim0.2\arcsec$ images obtained with the
   NACO spectrograph that night. In the end, none of the three
   previous efforts at reducing this Close et al. (2005) spectral
   dataset could obtain any meaningful K continuum of AB Dor C (and no
   H band spectra was attempted). So all these past efforts simply fit
   the "continuum" with a polynomial and attempted to remove as much
   of it as possible (in Close et al. 2005 the continuum was restored
   with a standard). Therefore, no spectrum of AB Dor C with the
   original continuum intact have been published to date.

   Without continua it is very hard to be sure that one is not either
   over or undersubtracting A from C. Hence, strengths of the CO and
   Na absorption lines are somewhat suspect and uncertain. This
   explains, in part, the significant dispersion in the spectral types
   determined from the three different reductions of the rather poor
   quality AO long slit spectra. To better understand what the real
   spectral type of AB Dor C is we have approached this problem with a
   new (more advanced) technique of PSD IFS reduction.

   Our IFS spectrum in this paper is superior for several reasons. One
   is that C was a more favorable $0.20\arcsec$ away from A at the
   time of the IFS observation. Another is that use of the IFS allowed
   significantly better removal of the contaminating PSF wing of A (at
   every wavelength) of the IFS datacube by PSF subtraction (see paper
   1 for full details). Moreover, we could easily check in each IFS
   "image" of how well we had removed the wing of A from the position
   of C (this is nearly impossible to do with long slit
   spectra). Also, by effectively integrating longer we were able to
   obtain S/N$\sim$40. In addition, we have obtained H band as well as
   K band spectra. Increasing the spectral range through H helps
   determine a more accurate spectral type. Finally, use of the the
   IFS eliminates slit losses, and has allowed us to produce excellent
   spectra of AB Dor C in the H and K bands with a stable continuum
   across both bands. To highlight the improvement we directly compare
   our new spectra with that of Nielsen et al. (2005) in figure 2 of
   paper 1.

   We believe that the points raised above make clear why our H \& K
   IFS continuum-preserved spectrum is superior to previously
   published (non-continuum preserved) K spectra of AB Dor C. However,
   it is still difficult obtaining a perfect spectral type fit for AB
   Dor C since it has an age (and so surface gravity) different from
   that of published M standards. For example, in Fig 5 it is clear
   that M4 is not the correct spectral type since the Na lines of Gl
   213 fit the AB Dor C spectra too well -- and since AB Dor C is younger
   than Gl 213 it should have {\it weaker} Na I lines than a field dwarf of
   the same spectral type (see Gorlova et al. 2004 for example). In
   fact, the M5 field dwarf Gl 51 fits the CO lines the best
   (including an exceptionally good fit to the pseudo-continuum past
   2.3 microns). Since CO is not strongly gravity dependent (Gorlova
   et al. 2004) there is evidence that the temperature of AB Dor C
   must be close to that of Gl 51 (M5). 

Spectra presented in previous
   studies all likely had too much strength in the Na I line compared
   to CO (which is unlikely for a low surface gravity object). In this
   study it is now clear that in the K band data AB Dor C appears as a
   low surface gravity M5 (fitting the CO well but weaker in the Na I
   lines). Moreover, our K-band continuum follows that of Gl 51 (M5)
   or Gl 406 equally well, yet that of an M4 (too hot) or M7 (too
   cold) can be excluded at $>1\sigma$ of our estimated noise level
   (see our $1\sigma$ errorbars in figures 5 \& 6). Hence, there is
   excellent evidence from Fig 5 that the spectral type of C must be
   in the range of M5-M6.

   Further evidence comes from Fig 6 (our H band spectra). Here we see
   the Mg \& Al lines between 1.65-1.75 microns of the M3 \& M4 dwarfs
   are not seen in the cooler AB Dor C atmosphere. Moreover, AB Dor
   C's continuum shape is nicely matched by the M5 and M6 templates
   from 1.58-1.8 microns. Whereas the bluest part of H has the lowest
   Strehl and likely some slight contamination from the hotter A
   component -- heating up our continuum slightly at the bluest part
   of H band.  The M7 continuum appears too cool. Hence, we see in the
   H band the best fit is to a M5-M6 spectral type. Therefore, we
   determine that the gravity independent line strengths (like CO) and
   the continuum of AB Dor C are best fit by field M5-M6 dwarfs. In
   summary, we conservatively estimate a spectral type of M$5.5\pm1.0$
   for AB Dor C.

\begin{figure}
 \includegraphics[angle=0,width=\columnwidth]{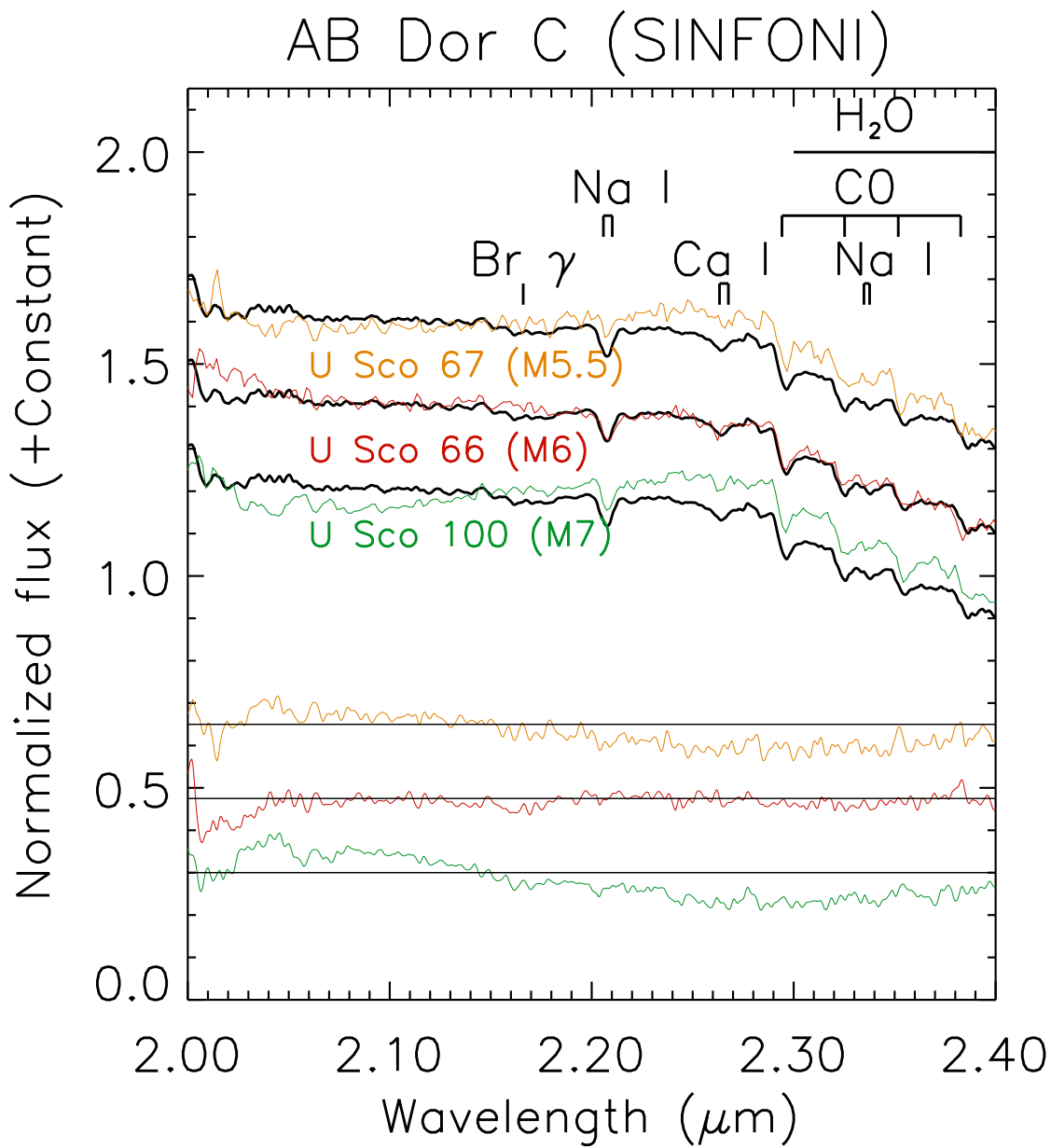}
\caption{
The $F_{\lambda}$ spectrum of AB Dor C (BLACK LINE -- smoothed to template resolution) shown against a number of young ($\sim$5 
Myr), 
low-surface gravity objects (colored lines; Gorlova et al. 2003).  The features seem to be similar to the M6 template. The bottom curves are the residuals from the 
spectrum of AB Dor C and the various templates. Note the rather low S/N of these faint young templates and small errors in the extinction correction leads to some uncertainty in their continuum which makes a good spectral type determination difficult, hence more weight should be placed on the higher S/N field M dwarfs shown in figures 5 \& 6. 
}
\label{fig4}
\end{figure}

\begin{figure}
\includegraphics[angle=0,width=\columnwidth]{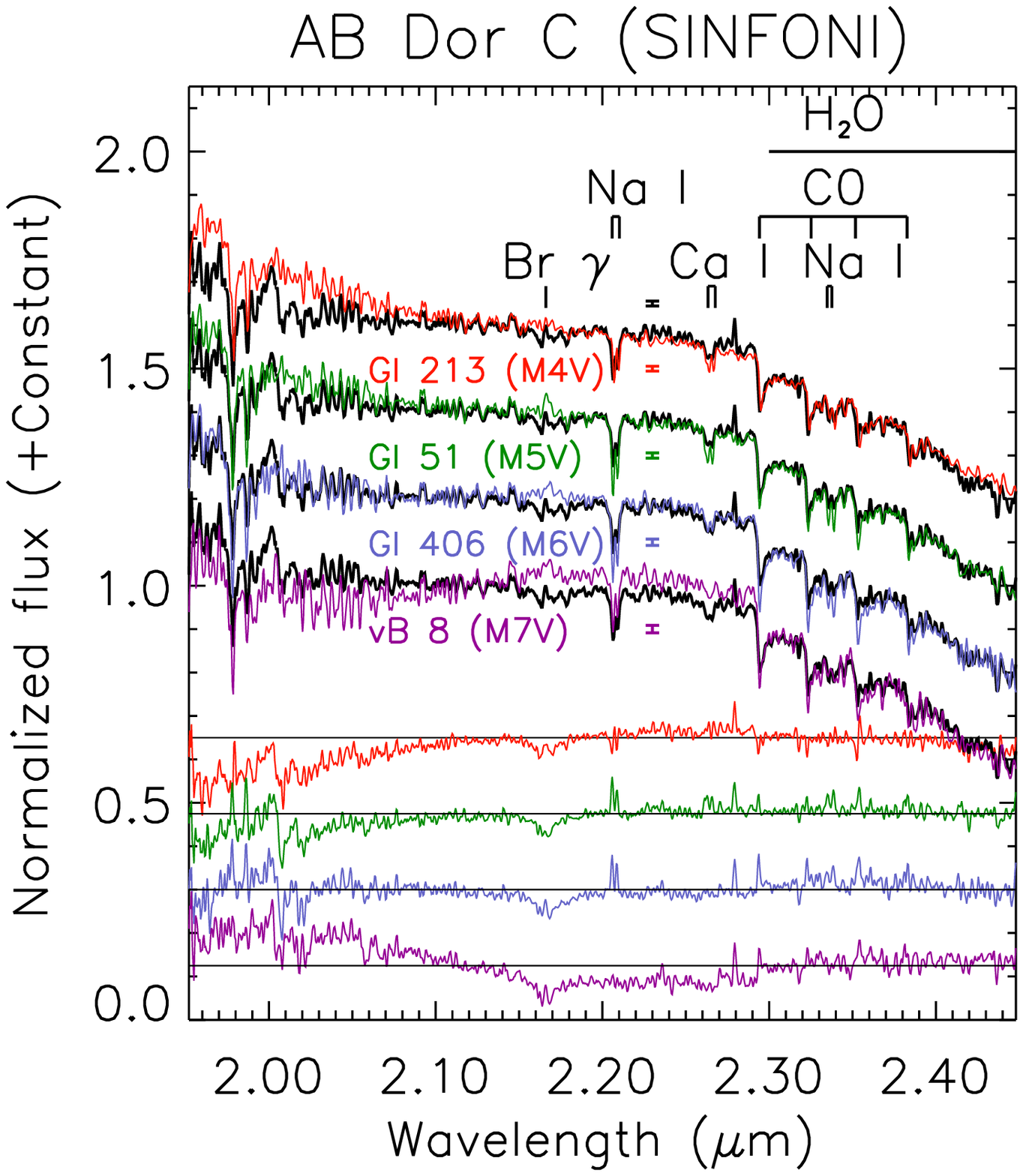}
\caption{
The unsmoothed $F_{\lambda}$ K spectrum of AB Dor C, this time plotted against high S/N field M dwarfs
($\sim$ 5 Gyr), with higher surface gravities (Cushing et al. 2005).
A spectral type of M5.5$\pm$1.0 (1$\sigma$) seems most consistent with
our spectrum since the gravity independent CO lines fit the M5 best but the continuum fits the M6 best. Note how the Na I lines are weaker than the M5 template but the CO fits very well, this is what we would expect for a young M5. The ``emmission'' line at 2.28 microns is due to a poor telluric subtraction.}
\label{fig5}
\end{figure}

\begin{figure}
\includegraphics[angle=0,width=\columnwidth]{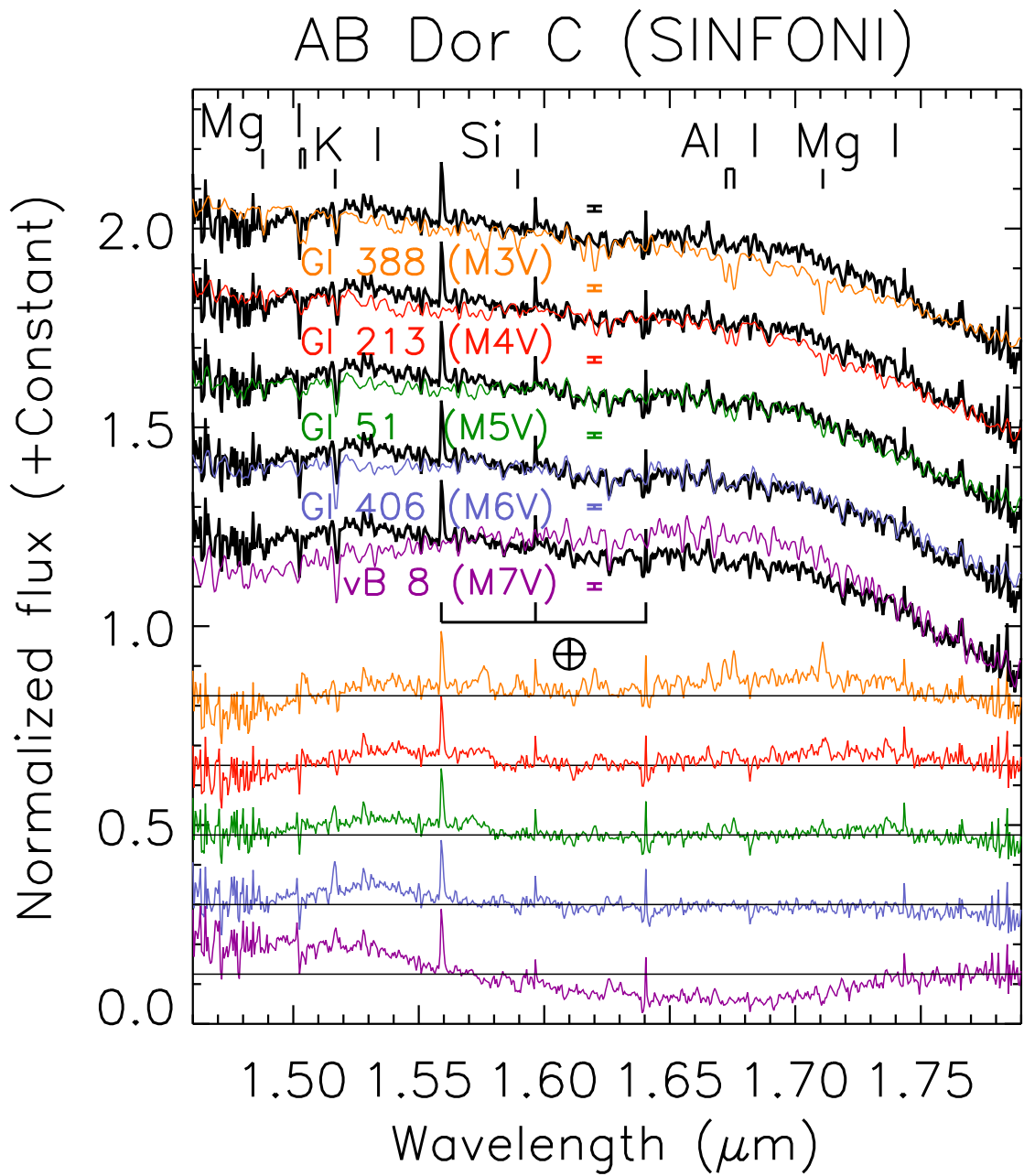}
\caption{
The unsmoothed $F_{\lambda}$ H spectrum of AB Dor C, plotted against field M dwarfs
($\sim$ 5 Gyr), with higher surface gravities (Cushing et al. 2005).
A spectral type of M5.5$\pm$1.0 (1$\sigma$) seems most consistent with
our spectrum since the Mg \& Al lines between 1.65-1.75 $\mu m$ fit the M6 slightly better but the 1.57-1.8 micron continuum fits the M5 or the M6 equally well. The ``emmission'' lines at 1.56, 1.64, and 1.74 microns are all due a poor telluric subtraction.  }
\label{fig6}
\end{figure}

\placefigure{fig4}
\placefigure{fig5}
\placefigure{fig6}

    Our new spectral type of M$5.5\pm1.0$ for AB Dor C is considerably
    earlier than the M$8\pm1$ measured in Close et al. (2005) and
    Nielsen et al. (2005) yet is consistent with the M$6\pm1$
    independently derived by Luhman \& Potter (2005). In summary, our continuum is preserved and the S/N ($\sim40$), wavelength
    range, and resolutions ($\sim$1500) are all much higher, hence the spectrum
    presented here is a significant improvement over
    any previous spectrum of AB Dor C.

Using the dwarf temperature scale of Leggett et al. (1996) we find a temperate range of $T_{eff} = 2925^{+170}_{-140}$ K for
AB Dor C. However, this error does not take into account a possible $\sim150$K systematic error in this popular temperature scale. Hence the final total errors could be as high as $2925^{+226}_{-205}$ K.

\section{Comparison to Models}

In figure 7 we compare our observed values to those predicted by the
cooling curves of the DUSTY models. Note how the J and H values appear
to be fainter than the models would predict (assuming a 75 Myr age
for AB Dor; Nielsen et al 2005), this result is similar to that of
Close et al. (2005).  Yet there is good agreement with the Ks (as was
also the case in Close et al. 2005). Moreover, our higher $T_{eff}$ is in
better agreement with the cooling curves than in Close et al. (2005).

\begin{figure}
\includegraphics[angle=0,width=\columnwidth]{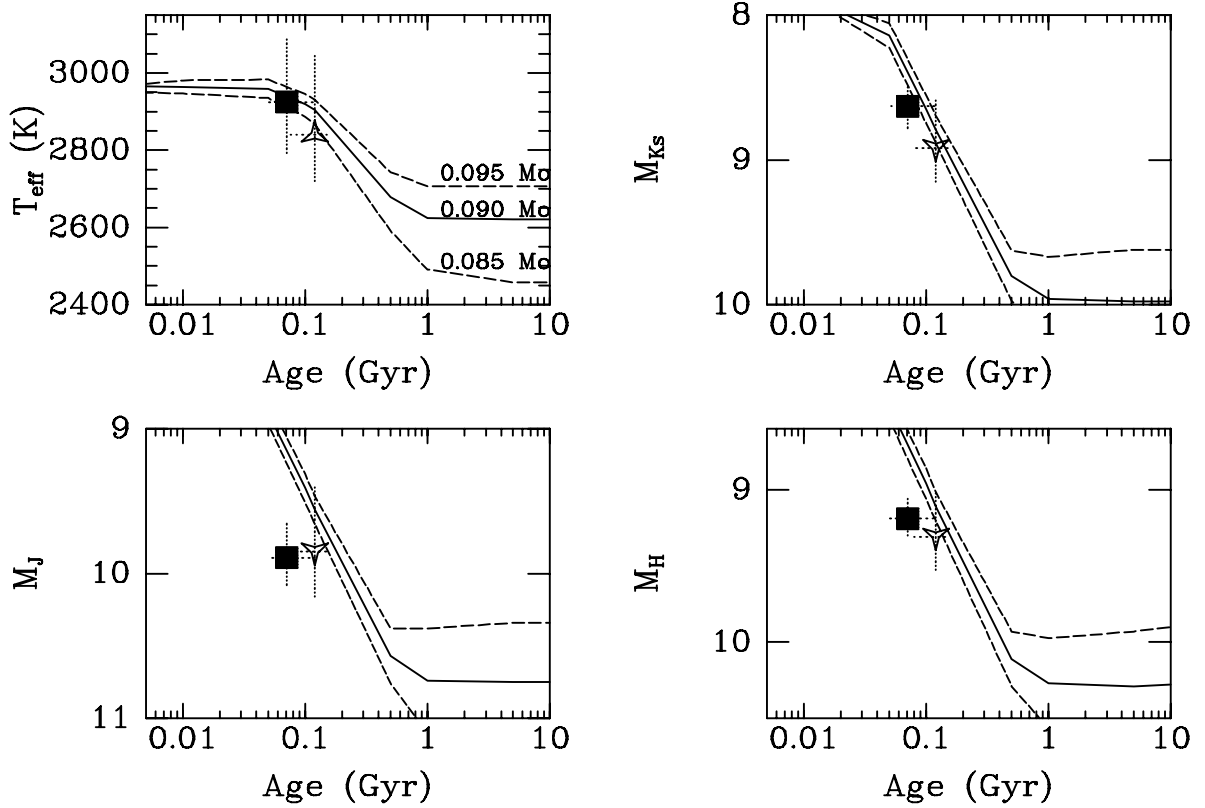}
\caption{
Comparison of our observations of AB Dor C (J\&H from Close et al.(2005); Ks and $T_{eff}$ from this work; all shown as solid squares; age 75 Myr) and those of Luhman \& Potter
(open triangles; plotted with their assumed older 120 Myr age) to the 0.095, 0.090, and 0.085 $M_{\sun}$ cooling
curves from the DUSTY models of Chabrier et al. (2000). Note the good
agreement with our new Ks and $T_{eff}$ observations; while there is
some overshoot in the J and H predicted fluxes for a 0.090
$M_{\sun}$ object.  }
\label{fig7}
\end{figure}

\placefigure{fig7}

It is interesting to note that the J-Ks for AB Dor C is
$\sim1.26\pm0.21$ which is surprisingly red for M5.5 object (which are
typically closer to $J-Ks\sim1.0$ in the Pleiades). Hence, it is
possible that the J band flux measured by Close et al. ($10.76\pm0.1$)
or Luhman \& Potter ($10.72^{+0.40}_{-0.63}$) may be systematically too
faint. We caution that it is problematic to obtain high-quality
high-contrast images in the J band with AO and so, in general, weight
should not be placed on J band fluxes of faint companions since lower
Strehls lead to a higher chance of a poor flux calibration. In any
case, the mean J flux of Close et al. (2005) would only have to be
brightened by $\sim0.2$ mag to give consistent colors with M5.5
Pleiades objects. This would lead to slightly better agreement with
the models especially at the oldest ages suggested for AB Dor of
75-150 Myr by Luhman et al. (2005)).

\begin{figure}
\includegraphics[angle=0,width=\columnwidth]{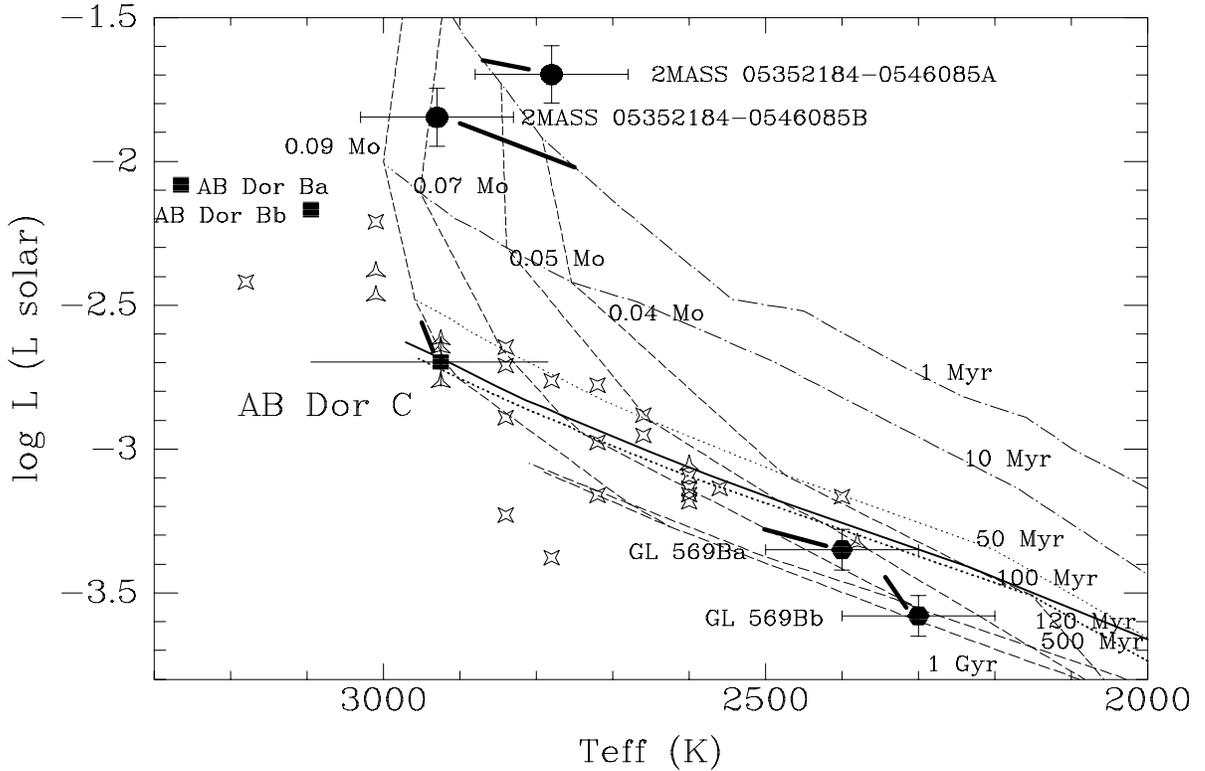}
\caption{
HR diagram showing low-mass Pleiades objects from Martin et al.  2000
(open stars), other low-mass members of the Pleiades taken from the
literature (open triangles), and AB Dor Ba/Bb (filled boxes).  The
dashed ``vertical'' lines are iso-mass contours for the DUSTY models (from
left to right: 0.09, 0.07,0.05, and 0.04 M$_{\sun}$), while the more
``horizontal'' lines are the DUSTY isochrones (top to bottom: 1, 10,
50, 100, 120, 500, 1000 Myr).  Note that the DUSTY models predict a
75 Myr object of 0.09 M$_{\sun}$ should be similar in temperature
and luminosity to that observed. From the location of AB Dor C on the HR diagram, one
would derive a mass of $\sim$0.09 M$_{\sun}$, a good estimate of the
measured mass.  As the temperature (dwarf $T_{eff}$ scale of Leggett et al. 1996; Luhman 1999) and luminosities ($BC_{K}$ of Allen et al. 2003) of the Pleiades
objects in this plot were determined in the same manner used for AB
Dor C, and these Pleiades points mostly fall along the appropriate 120
Myr DUSTY isochrone (dotted line), we are assured that our temperature
scale and bolometric correction are reasonable.  
The points with error bars mark all known low-mass young objects with well
determined dynamical masses, with the thick short diagonal lines
representing the displacement from the measured luminosity and
temperature to the values actually predicted by the DUSTY models. One
can see that for AB Dor C and and the older (300 Myr) Gl 569 B system
(Zapatero Osorio et al.  2004), the offset between observation and the
tracks are within the 1$\sigma$ uncertainties (especially considering
the uncertainty in the ages for these systems). There is somewhat less agreement for the secondary of the very young 2MASS 0535 eclipsing binary system in Orion (Stassun et al. 2006).} 
\label{fig8}
\end{figure}

\placefigure{fig8}

As is mentioned in Luhman et al. (2005) a good technique for
indirectly determining the mass of an object is to place the object in the HR
diagram and from the evolutionary tracks estimate a mass. In figure 8
we show how the new spectral type of M5.5 is in good agreement with
the DUSTY tracks in the HR diagram for a $0.090M_{\sun}$ object like
AB Dor C. Hence, the combination of a careful measurement of the Ks
flux and spectral type can allow the mass of a young low-mass object
like AB Dor C to be estimated from the DUSTY tracks. However,
underestimates of the mass can occur without proper (continuum
preserved) spectra or relying on just H and J fluxes alone.

The points with error bars in Figure 8 mark all low-mass young objects
with well determined dynamical masses. The thick diagonal lines
represent the rough disagreement between the measured luminosity and
temperature to the values actually predicted by the DUSTY models for
the true mass of the object. One can see that for AB Dor C and and the
older ($\sim$300 Myr) Gl 569 B system (Zapatero Osorio et al.  2004), the
offset between observation and the tracks are within the 1$\sigma$
uncertainties (especially considering the uncertainty in the ages for
these systems). 

However, in the recently discovered eclipsing binary 2MASS
J05352184-0546085 (Stassun et al. 2006) in the Orion Trapezium star
formation region ($\sim 1$ Myr age) the 0.034$\pm0.0027M_{\sun}$
secondary is {\it hotter} than the $0.0541\pm0.0046 M_{\sun}$ M6.5
primary. Hence, there is an increase in uncertainty in the models for
 very young ($>10$ Myr) low-mass ($<0.040 M_{\sun}$)
objects. However, the derived mass for 2M0535A would only be low by
$\sim25$\% (just consistant with the 1 Myr track), whereas the
derived mass for 2M0535B from the 1 Myr track would be about 200\% too
high (a similar trend was predicted by Mohanty et al. 2004).

\acknowledgements

We thank Mike Cushing and Nadja Gorlova for providing spectra of many young, 
low-mass objects. The Ks data was from observations made with the European Southern Observatory telescopes obtained from the ESO/ST-ECF Science Archive Facility under program ID 074.C-0084(B).

This publication makes use of data 
products from the Two Micron All Sky Survey, which is a joint project of 
the University of Massachusetts and the Infrared Processing and Analysis 
Center/California Institute of Technology, funded by the National 
Aeronautics and Space Administration and the National Science Foundation.

LMC is supported by an NSF CAREER award and the NASA Origins of Solar Systems program. ELN is supported by a Michelson Fellowship. NT, MT and FC are supported by a Marie Curie
Excellence Grant from the European Commission MEXT-CT-2003-002792
(SWIFT).

%% Tables should be submitted one per page, so put a \clearpage before

%% deluxetable environment provided by the AASTeX package or the LaTeX
%% table environment.  Use of deluxetable is preferred.
%%

%% Three table samples follow, two marked up in the deluxetable environment,
%% one marked up as a LaTeX table.

%% In this first example, note that the \tabletypesize{}
%% command has been used to reduce the font size of the table.
%% Note also that the \label command needs to be placed 

\clearpage
\begin{deluxetable}{lllll}
\tabletypesize{\scriptsize}
\tablecaption{The AB Dor System\label{tbl-0}}
\tablewidth{0pt}
\tablehead{
\colhead{Parameter} &
\colhead{A} &
\colhead{C} &
\colhead{Ba} &
\colhead{Bb} 
}
\startdata
K$_s$ magnitude & $4.686\pm0.016$ & $9.50\pm0.16$ & $8.08\pm0.20$ & $8.30\pm0.20$\\
Separation\tablenotemark{a}   to A & -- & $0.219\pm0.008\arcsec$ & $8.87\pm0.10\arcsec$ & -- \\
PA\tablenotemark{a}   w.r.t A & -- & $155.92\pm0.50^{o}$ & $346.31\pm0.50^{o}$ & -- \\
Separation\tablenotemark{a}  to Ba & -- & -- & -- & $0.060\pm0.003\arcsec$ \\
PA\tablenotemark{a}   w.r.t. Ba & -- & -- & -- & $246\pm2^{o}$ \\
Separation\tablenotemark{f}   to A & -- & $0.202\pm0.010\arcsec$ & -- & -- \\
PA\tablenotemark{f}   w.r.t A & -- & $180.78^{o}$ & -- & -- \\
Period w.r.t A (yr) & -- & $11.74\pm0.07$\tablenotemark{c} & 1400-4300\tablenotemark{b} & -- \\
Period w.r.t Ba (yr) & -- & -- & -- & $\sim0.9$ \\
Spectral Type & K1 & M$5.5\pm1.0$ & M$3.5\pm1.5$ & M$4.5\pm1.5$ \\
$T_{eff}$ (K) & $5081\pm50$ & $2925^{+170}_{-145}$ & $3265\pm245$ & $3095\pm255$ \\ 
Luminosity ($L_{\sun}$) & $0.388\pm0.008$ & $0.0021\pm0.0005$ & $0.008\pm0.002$ & $0.006\pm0.002$ \\ 
% = 10**((-1.0*((8.08-0.866+0.03+2.7)-4.74)/2.5)) calc of L_Ba
Mass ($M_{\sun}$)& $0.865\pm0.034$ & $0.090\pm0.005$\tablenotemark{c} & $<0.25$\tablenotemark{b} & $<0.15$\tablenotemark{b} \\
System Age (Myr) & $75\pm25$\tablenotemark{d} & & &\\
System Distance\tablenotemark{e} (pc) & $14.9\pm0.1$ & & &\\
\enddata
\tablenotetext{a}{observations made on Jan 7, 2005 UT (2005.0170)}
\tablenotetext{b}{Orbital calculations of Guirado et al. (2006)}
\tablenotetext{c}{Orbital solution of Nielsen et al. (2005)}
\tablenotetext{d}{Age solution of Nielsen et al. (2005)}
\tablenotetext{e}{Hipparcos distance ($15.06\pm0.11$ pc is derived by Hipparcos \& VLBI in Guirado et al. (2006))}
\tablenotetext{f}{observations made on Jan 24, 2006 UT (2006.0660)}
\end{deluxetable}

\clearpage

% FIGURES
%

%\begin{figure}
%\includegraphics[angle=0,width=\columnwidth]{f7.ps}
%\caption{
%The unsmoothed H \& K spectrum of AB Dor C (normalized at 1.6$\mu m$ in the H band only), plotted against field M dwarfs
%($\sim$ 5 Gyr), with higher surface gravities (Cushing et al. 2005).
%A spectral type of M5.5$\pm$0.5 (1$\sigma$) seems most consistent with
%our spectrum since the absorption lines fit the M6 but the continuum across H\&K fits the M5 best.  }
%\label{fig7}
%\end{figure}

%% The following command ends yourmanuscript. LaTeX will ignore any text 
\end{document}